\documentclass[pdflatex,sn-mathphys-num]{sn-jnl}% Math and Physical Sciences Numbered Reference Style
\usepackage{graphicx}%
\usepackage{multirow}%
\usepackage{amsmath,amssymb,amsfonts}%
\usepackage{amsthm}%
\usepackage{mathrsfs}%
\usepackage[title]{appendix}%
\usepackage{xcolor}%
\usepackage{textcomp}%
\usepackage{manyfoot}%
\usepackage{booktabs}%
\usepackage{algorithm}%
\usepackage{algorithmicx}%
\usepackage{algpseudocode}%
\usepackage{listings}%
\usepackage{physics}   
\usepackage{wasysym} 
\usepackage[
    starfontserif % comment for sans glyphs
    ]{starfont}
\usepackage{amsmath}
\theoremstyle{thmstyleone}%
\theoremstyle{thmstyletwo}%

\theoremstyle{thmstylethree}%

\begin{document}

\title[Article Title]{Dark matter heating of Planet 9, and its observational implications }
%- a possible observational signature of a distant object in the Solar System  }

%%=============================================================%%
%% GivenName	-> \fnm{Joergen W.}
%% Particle	-> \spfx{van der} -> surname prefix
%% FamilyName	-> \sur{Ploeg}
%% Suffix	-> \sfx{IV}
%% \author*[1,2]{\fnm{Joergen W.} \spfx{van der} \sur{Ploeg} 
%%  \sfx{IV}}\email{iauthor@gmail.com}
%%=============================================================%%

\author*[1,2]{\fnm{Tiberiu} \sur{Harko}}\email{tiberiu.harko@aira.astro.ro}

%\author[2,3]{\fnm{Second} \sur{Author}}\email{iiauthor@gmail.com}
%\equalcont{These authors contributed equally to this work.}

\affil*[1]{\orgdiv{Department of Physics}, \orgname{Babe\c s-Bolyai University}, \orgaddress{\street{1 Kog\u alniceanu Street}, \city{Cluj-Napoca}, \postcode{400083}, \state{Cluj}, \country{Romania}}}

\affil[2]{\orgdiv{Astronomical Observatory}, \orgname{Romanian Academy of Sciences}, \orgaddress{\street{19 Cire\c silor Street}, \city{Cluj-Napoca}, \postcode{400487}, \state{Cluj}, \country{Romania}}}

%%==================================%%
%% Sample for unstructured abstract %%
%%==================================%%

\abstract{The observed unusual behaviors of the orbits of Trans-Neptunian objects as well as the gravitational anomalies detected by the Optical Gravitational Lensing Experiment can be explained by assuming the existence of a ninth planet in the Solar System, having a mass of the order of $5-10M_{\earth}$, and located at the distance of 300-1000 AU from the Sun. However, since no optical counterpart of  Planet 9 was observed, it is reasonable to assume that it must have a very low luminosity. In this context various proposals on the nature of Planet 9 have been been advanced, including the possibility that it is a black hole, an axion or a dark matter star. In the present study we propose that dark matter heating of Planet 9 could generate a thermal radio flux that could allow its observational detection, even if Planet 9 is a very dark object. As a first step in this study we estimate the dark matter impact parameter, the mass and the kinetic energy deposition rates, as well as the surface temperature of Planet 9. By adopting a specific model for the time evolution of the planet, under the assumption of a long time capture of dark matter, the surface temperature of Planet 9, and the spectral features of the emitted radiation are obtained. Our results indicate that  dark matter capture may provide an efficient mechanism for the heating of Planet 9, and also provide a specific observational signature of the planet. The numerical evaluations depend  on the unknown value of the dark matter-ordinary matter interaction cross-section, with the estimates obtained as a function of its ratio and the saturation cross section for dark matter to deposit its entire energy. For    a value of this ratio of $10^{-10}$, and for a dark matter density of the order of $1.32\times 10^{-17}$ g/cm$^3$, in around Gyr the surface temperature of Planet 9 can reach values of the order of 200 K, or even higher, with a maximum wavelength of around $\lambda_{max}=1.44\times 10^{-3}$ cm, situated in the infrared domain. }

\keywords{Planet 9, Dark matter, Dark matter heating, Thermal radio flux}

%%\pacs[JEL Classification]{D8, H51}

%%\pacs[MSC Classification]{35A01, 65L10, 65L12, 65L20, 65L70}

\maketitle

\tableofcontents

\section{Introduction}\label{sec1}

Currently, there is mounting evidence that a massive planet, located well beyond the orbit of Neptune, may be present as the ninth planet in the Solar system. The first evidence for the existence of the planet came from the discovery of a population of eccentric Kuiper belt objects (KBOs), located in the outer Solar System, beyond  Neptune's orbit,  and decoupled from gravitational interactions with the planet \cite{Gom, Truj}. These various astronomical observations were explained in a unified framework in \cite{Brown1}, where it was recognized that the distant eccentric KBOs, moving outside the gravitational interaction with the planet Neptune, bunch together in the longitude of the perihelion. This means that there is an approximate alignment of their orbital axes. Moreover, the bunching together in the orbital plane of the orbits of the KBOs implies that their
angular momentum vectors are also aligned approximately, having 
similar values of the longitude of the
ascending node $\Omega$ and of inclination $i$, respectively \cite{Brown1}. 

Hence, the observed  clustering of Trans Neptunian Objects can be explained
by assuming the existence of a giant and yet unobserved planet, moving around the Sun on an inclined eccentric orbit \cite{Brown1, Brown2, Brown3}.  The presence in the Solar System of such a giant planet, called Planet 9, or P9,  could convincingly explain not only the oberved alignment of the orbital planes and axes of the distant KBOs, but also the large perihelion distances of objects like Sedna, a small dwarf planet located far beyond the orbit of Neptune, and moving around the Sun with a period of around 11,400 years \cite{Brown3}.

The orbital distribution and clustering of the distant KBOs is significantly influenced by the astronomical properties of Planet 9, like the mass and its orbital elements.  Thus the astronomical observations of the KBOs can be used to infer the unknown astronomical parameters characterizing the Planet 9, including its mass and its orbital elements. A Markov Chain Monte Carlo analysis performed in \cite{Brown4} allowed to estimate the physical and astronomical parameters of Planet 9, giving for the planet a mass of  $6.2^{+2.2}_{-1.3}M_{\oplus}$,  a semimajor axis of  $380^{+140}_{-80}$ AU, a perihelion of the order of $300^{+85}_{-60}$ AU, and an inclination angle of  $16 \pm  5^{\circ}$, respectively.

  In \cite{Khain} it was shown that although some resonant interactions in the orbits of the  Trans-Neptunian Objects (TNOs),  small astronomical bodies located in the distant Kuiper belt,  with Planet 9 do exist, the anti-aligned astronomical objects can survive without the resonances. This confirms the result that the dynamics of the TNOs are essentially determined by secular, rather than resonant, interactions with P9.

  The hypothesis of the existence of Planet 9 has led to extensive observational searches for the planet. A new custom-designed pipeline  created to search for very low luminosity, yet undiscovered Solar System bodies by using full-frame image data from the NASA Transiting Exoplanet Survey Satellite (TESS) mission was presented in \cite{obs1}. Without the need of any orbital information, the pipeline can accurately recover the electromagnetic signals of Solar System bodies located in the Galactic plane far away from the Sun for distances $d \lesssim 150$ AU. 
  
  In \cite{obs2} it was shown that P9 can induce dynamical evolutions  leading to the orbital evolution in the inner Oort cloud, a dynamically frozen system. Hence  astronomical objects located in the inner Oort cloud,  can obtain orbital characteristic from the distant scattered disk.  This implies that the observed statistics of the long-period TNOs is represented by a mixture of Kuiper Belt and Oort cloud objects.

   The possibility of the six years observations of the Dark Energy Survey (DES) to retrieve a synthetic Planet Nine population was investigated in \cite{obs3}. In \cite{obs4} it was proposed that the study of the orbital elements of some meteoroids arriving at Earth could have the potential of indicating the presence of a massive trans-Neptunian planet. A possible candidate that may offer some clues for the existence of Planet 9 may be the meteor CNEOS 2014-01-08, a meteor that may be the first observed interstellar object of this type, due to ts hyperbolic orbit. As a possible candidate location for Planet Nine the region of coordinates R.A. $53.^{\circ} 0 \pm 4.^{\circ}3$, and decl. $9.^{\circ}2 \pm 1.^{\circ}3$, respectively,  was also proposed.

   A search for Planet Nine using the second data release of the Pan-STARRS1 (PS1) survey was presented in \cite{obs5}.  The existence of a Planet Nine with the characteristics in \cite{Brown4} was ruled out to a 50\% completion depth of $V = 21.5$. Moreover, the survey, along with previous analyses of the Zwicky Transient Facility (ZTF) and Dark Energy Survey (DES) data, rules out at a level of 78\% of the \cite{Brown4} parameter space. The combination of the ZTF, DES, and PS1 surveys rules out 78\% of the \cite{Brown4} reference population. 
   
   The updated estimates for Planet 9 \cite{obs5} predict a mass of $6.6^{+2.6}_{-1.7} M_{\oplus}$, a semimajor axis of $500^{+170}_{-120}$ AU,  an aphelion distance of $630^{+290}_{-170}$ AU, a current distance of $550^{+250}_{-180}$ AU, and a V magnitude of $22.0^{+1.1}_{-1.4}$, respectively.
   
Comprehensive N-body simulations that self-consistently model gravitational perturbations from all giant planets, passing stars, as well as the Galactic tide, by using initial conditions that consider the primordial migration of the giant planets, as well as the Sun's early evolution within a star cluster, were performed in \cite{obs6}. The simulation results show that the orbital characteristics of exotic orbits, ranging from those with high perihelia to those with extreme inclinations, can be explained by the Planet 9 hypothesis, and an inclusive model for the existence of the giant planet. 

A search for Planet Nine in the far-infrared, where the peak of the black body radiation is located, using the results of the most sensitive all-sky far-infrared survey realized to date, AKARI, was carried through in \cite{obs7}. While in the optical searches the energy of the reflected electromagnetic radiation (sunlight) decreases according to the  $d^4$ law, in the infrared domain the thermal radiation decreases with the square of the  distance $d$ to the Sun, as $d^2$.  The investigation performed in \cite{obs7} indicated the possible existence of two  Planet Nine candidates, whose positions and fluxes are in the theoretical prediction domains.

Planet Nine candidates were investigated by using two far-infrared all-sky surveys,  IRAS and AKARI, respectively, in \cite{obs8}. These two surveys were separated by a period of 23 years, a time interval that may allow to detect Planet Nine's predicted  orbital motion of $\sim 3'$/year. For the search of the planet a custom built AKARI Far-Infrared point source list was used. One good planet candidate was found, for which after 23 years the IRAS source can not be seen in the same coordinate as in the AKARI image, and conversely. However, it is important to point out that AKARI and IRAS observations alone are not enough to determine the full orbital characteristics of this candidate planet. 

A possible observational signature for Planet 9 was suggested in \cite{Chan}. The probability of capturing large Trans-Neptunian objects by the massive Planet Nine to form a satellite system in the scattered disk region (between the inner Oort Clouds and the Kuiper Belt) can be estimated as being rather large. Therefore the tidal effect of the planet can heat up the satellites significantly. The temperature increase of the satellites can give an enough high thermal radio flux for the observational detection of the planet, even if Planet 9 is a very dark object, with an extremely low electromagnetic emissivity. 

The physical nature of Planet 9 is also a matter of intense debate. In \cite{BH2}  it was suggested that the anomalous orbits of the TNOs and an excess in microlensing events in the 5-year OGLE (Optical Gravitational Lensing Experiment) dataset can be simultaneously explained by a assuming the existence of a new population of astrophysical objects having masses exceeding several times that of the Earth. These astrophysical objects were assumed to be primordial black holes (PBHs). If it happened that one of these PBHs was seized by the Solar System, the orbits of TNOs would be modified,  similarly to the P9 proposal. Hence, P9 may be a primordial black hole \cite{BH2,t1}. 

Due to their interactions with ordinary baryonic matter, non-annihilating dark matter particles can concentrate inside celestial objects \cite{t2}. If the dark matter particles have high masses, they are attracted gravitationally towards the core of the host planet, and finally they thermalize in a small region of the core. Eventually, via core collapse,  they can form small black holes. However, in this process the celestial object that captured dark matter is destroyed in the gravitational collapse process. 

 In \cite{t3} it was suggested that Planet 9 can be an axion star, consisting of clusters of condensed QCD axions or axionlike particles (ALPs), gravitationally bounded. The probability of the Solar System of capturing an axion star is of the same order of magnitude as the probability of capturing a free floating planet (FFP). This probability is even higher for the case of an axion star. Even though due to two-photon decay, axion stars can emit monochromatic electromagnetic signals, the frequency of the decayed photons is either not within the frequency range of the present day radio telescopes, or the decay signal is too weak to be detected. 

Modified Newtonian Dynamics (MOND), an alternative gravitational theory, which can explain the behavior of the galactic rotation curves without invoking the presence of dark matter \cite{MOND}, can also  provide an  explanation for the observed astronomical anomalies in the Solar System \cite{t4}.  Using the secular approximation in the MOND framework,  it was shown in \cite{t4} that the major axes of the orbits of the small objects align with the direction toward the Galactic center.  Moreover, the orbits cluster in the phase space, in agreement with observations of the TNOs. Hence, the effects of MOND could also be relevant and detectable in the outer Solar System.

   Dark matter represents the dominant matter component of the Universe, representing around 25\% of its matter-energy content.  Yet despite many observational and experimental searches \cite{s1,s2,s3,s3a,s3b}, the physical nature of dark matter remains elusive, thus motivating further astrophysical and cosmological investigations. While some dark matter models have been eliminated by experiments, many viable  dark matter models do exists \cite{s4}. 

   One of the interesting physical effects related to the presence of dark matter around stellar type objects is the possibility that the interactions between dark matter particles and ordinary (baryonic) matter can heat the latter through the deposition of the kinetic energy that dark matter gains due to its fall in the gravitational potential of a massive object \cite{h1,h2}. The dark kinetic heating of baryonic objects  depends on the total mass of the accumulated dark matter. Moreover, the kinetic heating by dark matter is also very easily affected to the mass range of the dark matter particle. Hence, if detected observationally, dark kinetic heating of baryonic matter made astrophysical objects can provide a strong evidence for the existence of dark matter. Moreover, it can also give a clear observational evidence for the presence of dark matter near or inside massive astrophysical objects.

It is the main goal of the present work to investigate the possibility of the dark kinetic heating of Planet 9, and the observational signatures such a heating may provide. As it has been already pointed out in \cite{h2}, dark kinetic heating can be important even in objects with low-escape velocities, and small gravitational potentials, such as exoplanets and brown dwarfs. Thus, dark kinetic heating may significantly increase the possibility of the discovery of such astrophysical objects. 

In our analysis we estimate the impact parameter of dark matter for Planet 9, and obtain the mass rate of dark matter crossing the planet, as well as the kinetic energy deposition rate. The knowledge of these quantities allows to estimate the surface temperature of the planet, by considering both a static scenario, as well a dynamic evolution model, in which the mass of the planet increases due to dark matter capture. The spectral characteristics of the black body radiation emitted by the planet surface are also briefly discussed.    

The present paper is organized as follows. We review the basic mechanisms of dark matter heating in Section~\ref{sec2}. The astrophysical and observational parameters of Planet 9, due to kinetic heating by dark matter, are considered in Section~\ref{sec3}. A time-dependent heating scenario is also presented. Finally, we discuss and conclude our results in Section~\ref{sec4}. 
   
\section{Dark kinetic heating of exoplanets}\label{sec2}

In the present Section we investigate the basic physical mechanisms that contribute to the kinetic heating of baryonic astrophysical objects by dark matter. The first fundamental quantity necessary for the estimation of the dark kinetic heating is the dark matter flux through the baryonic object. The flux essentially depends on the maximum impact parameter of the incoming dark matter. With the help of the impact parameter we obtain the expressions of the mass and kinetic energy deposition,  and we estimate the surface temperature, luminosity and spectrum of the planet.

The kinetic effects of dark matter on the cosmic objects essentially depend on two parameters: the local density of dark matter, and on its velocity.  Purely gravitational constraints on the density of dark matter  in the Solar System were obtained in \cite{dmd}, by using the tracking data of asteroid (101955) Bennu. This asteroid has extensive tracking data and high-fidelity trajectory modeling resulting from the OSIRIS-REx mission. According to this investigation the local
density of dark matter is bound by $\rho_\chi\lesssim 10^{-18}$ g/cm$^3$
  a value estimated around 1.1 AU. This constraint applies
to all dark matter candidates but are particularly including solar halos,
stellar basins, and axion miniclusters, which predict or allow overdensities in the Solar System. 

In the following we assume that the dark matter density distribution in the Solar System is at least approximately homogeneous, and we adopt the same value for $\rho_\chi$ in the  vicinity of Planet 9 as estimated in \cite{dmd} for the local density of dark matter. As for the velocity $v_\chi$ of dark matter, we adopt the value $v_\chi=250\;{\rm km/s}=2.50\times 10^7\; {\rm cm/s}$ \cite{Pato}, which gives a dark matter flux $j_\chi$ crossing Planet 9 of the order
\begin{equation}
j_\chi=\rho_\chi v_\chi=2.5\times 10^{-11}\times \left(\frac{\rho_\chi}{10^{-18}\;{\rm g/cm^3}}\right)\times \left(\frac{v_\chi}{2.50\times 10^7\;{\rm cm/s}}\right)\;{\rm g/cm^2\;s}.
\end{equation}

\subsection{The impact parameter of dark matter for P9}

In the following we will consider, for the sake of generality, a general relativistic description of the dark matter-ordinary matter interaction, by assuming that the metric around the compact astrophysical object, a planet in our case, is described by the Schwarzschild line element, $ds^{2}=e^{\nu (r)}c^{2}dt^{2}-e^{\lambda (r)}dr^{2}-r^{2}\left( d\theta ^{2}+\sin ^{2}\theta d\phi ^{2}\right) $, where the metric tensor components $\nu (r)$ and $\lambda (r)$ depend on the radial coordinate $r$ only. 

The motion in the Schwarzschild metric is geodesic, and there are two
conserved quantities, the energy $E_{\infty }=\left( 1-v_{\chi
}^{2}/c^{2}\right) ^{-1/2}c^{2}$ and the angular momentum $L=v_{\chi }b$ of the dark matter particle, respectively, where $v_{\chi }$ is the velocity, and $b$ is the impact parameter \cite{Gold}. 

For a particle moving very close to the surface of planet
of mass $M$ and radius $R$ from the geodesic equation of motion we obtain
the relation \cite{Gold}
\begin{equation}
\frac{\left( E_{\infty }/c^{2}\right) ^{2}}{1-2GM/c^{2}R}-\frac{L^{2}}{%
c^{2}R^{2}}=1.
\end{equation}

By assuming $v_{\chi }^{2}/c^{2}<<1$,  the impact parameter is given by the expression \cite{Gold}
\begin{equation}
b=\sqrt{r_{g}R\frac{c^{2}}{v_{\chi }^{2}}}\frac{1}{\sqrt{1-r_{g}/R}}=R\frac{%
v_{es}}{v_{\chi }}\left( 1-\frac{r_{g}}{R}\right) ^{-1/2},
\end{equation}
where $r_{g}=2GM/c^{2}$ is the gravitational radius of the planet, and $%
v_{es}=\sqrt{2GM/R}$ is the escape velocity of a particle from the surface of the planet. 

In the following for qualitative estimations for Planet 9 we assume a mass of $M=10M_{\oplus}$  and a radius $R=10R_{\oplus}$, where $M_{\oplus}=5.97\times 10^{27}$ g is the mass of the Earth, and $R_{\oplus}=6.371\times 10^{8}$ cm is the Earth's radius. This gives for the gravitational radius of the Planet 9 the value $r_{g}\approx
8.85$ cm, and thus $r_{g}/R<<1$. Hence for the impact parameter we obtain the expression 
\begin{equation}
b\approx \frac{\sqrt{2GMR}}{v_{\chi }}=2.85\times 10^{3}\times \left( \frac{M%
}{10M_{\oplus}}\right) ^{1/2}\times \left( \frac{R}{10R_{\oplus}}\right)
^{1/2}\times \left( \frac{v_{\chi }}{2.50\times 10^7\;{\rm cm/s}}\right) ^{-1}\;{\rm km},
\end{equation}
where for the velocity of the dark matter particle we have adopted a value of the order of 250 km/s \cite{Pato}. 

\subsection{Dark kinetic energy deposition}

The total mass rate of dark matter passing through the planet is defined as \cite{h1}
\begin{equation}\label{mass}
\frac{dm}{dt}=\dot{m}=\pi b^2j_\chi=\pi b^{2}\rho _{\chi }v_{\chi }\approx 2\pi GMR\frac{\rho
_{\chi }}{v_{\chi }},
\end{equation}%
where $\rho _{\chi }$ is the environmental density of dark matter.  For the case of Planet 9 we obtain the dark matter mass rate crossing the planet as
\begin{eqnarray}
\frac{dm_{9}}{dt}= 6.37\times 10^6&\times& \left( \frac{M}{%
10M_{\oplus}}\right) \times \left( \frac{R}{10R_{\oplus}}\right) \times \left( 
\frac{\rho _{\chi }}{10^{-18}\;{\rm g/cm^{3}}}\right) \nonumber\\
&&\times \left( \frac{v_{\chi }}{%
2.50\times 10^7\;{\rm cm/s}}\right) ^{-1}\;{\rm g/s}.
\end{eqnarray}

For the density of dark matter around Planet 9 we assume a value of the order of $\rho_\chi =10^{-18}\; {\rm g/cm^3}$ \cite{dmd}.

The total kinetic energy that can be deposited by dark matter on the planet can be approximated by the dark matter's kinetic energy at the surface of the planet \cite{h1}, which is given by $E_{ps}=\left( \gamma -1\right) m_{\chi }c^{2}$,  where $\gamma $ is the Lorentz factor of the dark matter $\gamma =\left( 1-v_{\varkappa
}^{2}/c^{2}\right) ^{-1/2}$. However, in the non-relativistic
approximation of the present approach we can write $E_{ps}=m_{\chi }v_{\chi}^{2}/2$. Then the rate of the kinetic energy deposition is given by \cite{h1}
\begin{equation}\label{Ek}
\frac{dE_{k}}{dt}=\frac{\dot{m}}{m_{\chi }}E_{ps}f=\pi GMRj_\chi f=\pi GMR\rho _{\chi
}v_{\chi }f,
\end{equation}%
where $m_{\chi }$ is the mass of the dark matter particle, and $f\in \left[0,1\right] $ is the fraction of dark particles crossing through the planet that become trapped in its interior. $f$  depends on the
scattering cross section of dark matter on nucleons and electrons. 

The
capture probability can also be written as 
$f\equiv {\rm Min}\left[ \sigma _{n\chi }/\sigma _{sat},1\right] $ \cite{h1}, where $\sigma_{sat}\left( m_{\chi }\right) $ is the cross section for which all dark matter is captured by the planet. 

 For the particular case of Planet 9 the kinetic energy deposition rate by dark matter is given by
\begin{eqnarray}
\frac{dE_{k9}}{dt}=1.99\times 10^{21}&\times&
\left( \frac{M}{10M_{\oplus}}\right) \times \left( \frac{R}{10R_{\oplus}}\right)
\times \left( \frac{\rho _{\chi }}{10^{-18}\;{\rm g/cm^{3}}}\right) \nonumber\\
&&\times \left( 
\frac{v_{\chi }}{2.50\times 10^7\;{\rm cm/s}}\right) \times \frac{f}{1}\;{\rm erg/s}.
\end{eqnarray}

 In order to estimate the range of values of the capture fraction $f$ we need first to evaluate the saturation cross-section, $\sigma_{sat}$, for which all incoming dark matter is captured. We assume in the following that dark matter is captured after it scatters once with the nucleons inside the planet, and it becomes bound after one single scattering event. Therefore  the saturation cross-section for dark matter to deposit all of its kinetic energy into the planet is the per-neutron cross section for which dark matter scatters once, and which is given by \cite{h1}
\begin{equation}
\sigma _{sat}\approx \pi R^2\frac{m_n}{M}=3.57\times 10^{-33}\times \left(\frac{R}{10R_\oplus}\right)^2\times \left(\frac{M}{10M_{\oplus}}\right)^{-1}\;{\rm cm^2}.
\end{equation}

 There are several observational/experimental constraints  for the dark matter-baryon interaction cross section and the dark matter particle mass for velocity-independent interactions  \cite{C1,C2,C3,C4, CS}. Experimental limits on the dark matter-nucleon cross section $\sigma_{n\chi}$ were obtained in the XENON \cite{C1}  and CDMS \cite{C2} experiments, which searched for energy deposition via nuclear recoils due to dark matter scattering. For the dark matter mass in the range 20 GeV and 100 GeV, the obtained limit is $\sigma_ {n\chi}<6\times 10^{-44}$ cm$^2$. For a larger dark matter particle mass, of the order of $m_\chi=103$ GeV, the limit is about an order of magnitude larger, and it is similar to dark matter particle masses of  $m_\chi\sim 15$ GeV. The CRESST-II experiment has obtained a 4.7$\sigma$ limit, corresponding to $m_\chi\sim 10-30$ GeV, and $10^{-40}\;{\rm cm^2}<\sigma _{n\chi}<3\times 10^{-43}\;{\rm cm^2}$ \cite{C3}. The LUX-ZEPLIN experiment, a dark matter detector centered on a dual-phase xenon time projection chamber, has obtained, for spin-independent scattering at 36 GeV/c$^2$ an upper bound of $ 9.2\times 10^{-48}\;{\rm cm^2}$ at the 90\% confidence level for dark matter - nucleon, neutron and proton scatterings. These experimental results suggest the approximate range $\sigma _{n\chi}\in \left(10^{-47}\;{\rm cm^2}, 10^{-40}\;{\rm cm^2}\right)$ for the dark matter-normal matter cross section. 

 Assuming the same interaction cross section of dark matter with all baryons, the range for the dark matter capture rate of the planet $f=\sigma_{n\chi}/\sigma_{sat}$ can be obtained as 
 \begin{equation}
f\in \left(2.79\times 10^{-15},2.79\times 10^{-8}\right).
 \end{equation}

\subsubsection{Dynamical evolution of the planet}

We consider now the possibility of the dynamical evolution of the planet due to the trapping of dark matter. The capture of the dark matter leads to an increase of the total mass $M$.    We assume that the rate of the mass increase can be obtained according to the definition
\begin{equation}
\frac{dM}{dt}=\alpha \frac{dm}{dt},
\end{equation}
where the dimensionless constant parameter $\alpha = (dM/dt)/(dm/dt)$ describes the rate of the mass increase. With the use of Eq.~(\ref{mass}) we obtain for the total mass variation the relation
\begin{equation}
\frac{dM(t)}{dt}=2\pi G\alpha R(t)M(t)\frac{\rho_\chi}{v_\chi},
\end{equation}
which can be integrated to give
\begin{equation}
M(t)=M_0\exp \left[2\pi G \alpha \frac{\rho_\chi}{v_\chi}\int_{t_0}^t{R(t)dt}\right],
\end{equation}
where $M_0=M\left(t_0\right)$ is the initial mass of the planet. With the use of the mean value theorem for integrals we can write $\int_{t_0}^t{R(t)dt}=R\left(t-t_0\right)$, where $R$ is the average value of the radius of the planet. Furthermore, we assume $t>>t_0$, which gives $\int_{t_0}^t{R(t)dt}\approx Rt$. Therefore we obtain for the mass variation due to dark matter capture of the planet the relation
\begin{equation}
M(t)=M_0e^{t/t_c}=M_0\exp\left[2\pi \alpha GR\frac{\rho_\chi}{v_\chi}t\right]=M_0e^{t/t_c},
\end{equation}
where
\begin{equation}\label{tc1}
t_c=\frac{v_\chi}{2\pi \alpha GR\rho_\chi},
\end{equation}
and
\begin{equation}\label{tc2}
t_c=9.36\times 10^{21}\times \frac{1}{\alpha}\times \left(\frac{v_\chi}{2.50\times 10^7\;{\rm cm/s}}\right)\times \left(\frac{R}{10R_\oplus}\right)^{-1}\times \left(\frac{\rho_\chi}{10^{-18}\;{\rm g/cm^3}}\right)^{-1}\;{\rm s}.
\end{equation}

For $t/t_c<<1$, we have 
\begin{equation}\label{mass1}
M(t)=M_{0}e^{t/t_{c}}\approx M_0\left(1+\frac{t}{t_c}\right), t/t_c<<1.
\end{equation}

%\begin{equation}
%M(t)=M_{0}\left[ 1+\alpha m(t)\right] ,
%\end{equation}%
%where $M_{0}$, representing the initial baryonic mass of the planet,  and %$\alpha $ are constants. Moreover, we impose the condition $m(0)=0$, and we %also assume that the radius $R$ of the planet does not change significantly, %and thus we can neglect its time variation. Then the mass rate equation of %the dark matter takes the form
%\begin{equation}
%\frac{dm(t)}{dt}=2\pi GM_{0}\left[ 1+\alpha m(t)\right] R\frac{\rho _{\chi }%
%}{v_{\chi }},
%\end{equation}%
%and it has the general solution 
%\begin{equation}
%m(t)=\frac{1}{\alpha }\left[ e^{t/t_{c}}-1\right] ,
%\end{equation}%
%where we have denoted
%\begin{equation}\label{tc}
%t_{c}=\frac{v_{\chi }}{2\pi \alpha GM_{0}R\rho _{\chi }}.
%\end{equation}

%Hence for the law of the variation of the mass of the planet due to dark %matter capture we obtain the relation
%\begin{equation}\label{mass1}
%M(t)=M_{0}e^{t/t_{c}}\approx M_0\left(1+\frac{t}{t_c}\right), t/t_c<<1.
%\end{equation}

By estimating the mass variation equation (\ref{mass1}) at the present time $t=t_{f}$, we obtain the relation
\begin{equation}
\frac{t_{f}}{t_{c}}=\ln \frac{M\left( t_{f}\right) }{M_{0}}=\frac{2\pi
\alpha GR\rho _{\chi }t_{f}}{v_{\chi }},
\end{equation}
which gives for $\alpha $ the expression
\begin{equation}
\alpha =\frac{v_{\chi }\ln \left[ M\left( t_{f}\right) /M_{0}\right] }{2\pi
GR\rho _{\chi }t_{f}},
\end{equation}
or
\begin{eqnarray}\label{alpha}
\alpha =6.60\times 10^4&\times &\left(\frac{v_\chi}{2.50\times 10^7\;{\rm cm/s}}\right)\times \left(\frac{R}{10R_\oplus}\right)^{-1}\times \left(\frac{\rho_\chi}{10^{-18}\;{\rm g/cm^3}}\right)^{-1}\nonumber\\
&\times& \left(\frac{t_f}{1.41\times 10^{17}\;{\rm s}}\right)^{-1}\times \ln\frac{M\left(t_f\right)}{M_0}.
\end{eqnarray}

For an initial planetary mass $M_{0}=1\times M_{\oplus}$, and a final mass $%
M\left( t_{f}\right) =10\times M_{\oplus}$, by assuming a time interval of 4.5
billion years, we obtain $\alpha =1.52\times 10^5$. Hence, in the following qualitative numerical estimations we will adopt for $\alpha $ the value $\alpha =1.50\times 10^5$.

As for the time variation of the rate of the kinetic energy deposition by
dark matter, it takes the simple form  
\begin{equation}
\frac{dE_{k}(t)}{dt}=\pi GM_{0}R\rho _{\chi }v_{\chi }e^{t/t_{c}}f\approx \pi GM_{0}R\rho _{\chi }v_{\chi }\left(1+\frac{t}{t_c}\right)f,
\end{equation}
giving 
\begin{equation}
E_k(t)=\frac{1}{\alpha} \frac{M_0v_\chi^2}{2}f\left(e^{t/t_c}-1\right),
\end{equation}
where we have used the initial condition $E_k(0)=0$.

\section{Dark matter heating of Planet 9}\label{sec3}

When a dark matter particle crosses the  surface of the planets, it is
scattered by the baryons (nuclei), and thus the kinetic energy of the dark matter is transferred to the planet. This leads to the heating of the planet, and to an increase in its surface temperature 

\subsection{Surface temperature of Planet 9}

After the thermalization of the scattered nuclei inside the planet, 
the  dark kinetic heating process induces a planet luminosity at large
distances given by \cite{h1}
\begin{equation}
L_{\infty }=\frac{dE_{k}}{dt}=\pi GMR\rho _{\chi }v_{\chi }f=4\pi \sigma _{B}R^{2}T_{S}^{4},
\end{equation}%
where $\sigma _{B}=5.67\times 10^{-5}$ erg/cm$^2$ K$^4$ is  the Stefan-Boltzmann constant, and $T_{S}$ is the
blackbody temperature of the planet's surface. 

In the static case, by using
the expression (\ref{Ek}) of the kinetic energy deposition by dark matter, we obtain for the surface temperature of the planet the expression
\begin{equation}
T_{S}=\left( \frac{2GM}{R}\frac{1}{8\sigma _{B}}\rho _{\chi }v_{\chi
}f\right) ^{1/4}=\left( \frac{1}{8\sigma _{B}}v_{es}^{2}\rho _{\chi }v_{\chi
}f\right) ^{1/4}=\left( \frac{1}{8\sigma _{B}}v_{es}^{2}j _{\chi }f\right) ^{1/4}.
\end{equation}

\begin{figure}[ht!]
    \centering
    \includegraphics[width=0.85\textwidth]{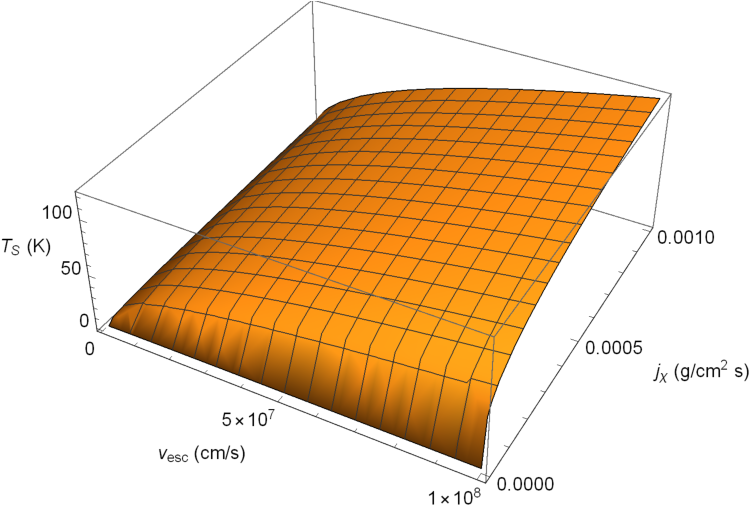}
    \caption{Variation of the surface temperature $T_S$ of Planet 9 as a function of $v_{esc}$ and $j_\chi$, for $f=10^{-8}$. }
    \label{fig1}
\end{figure}

The variation of $T_s$ as a function of $v_{esc}$ and $j_\chi$ is represented in Fig.~\ref{fig1}.

In Fig.~\ref{fig1} we have considered an escape velocity that ranges from its Earth value $v_{esc}\approx 10^6 \;{\rm cm/s}$ to an upper limit of $v_{esc}\approx 10^8\;{\rm cm/s}$. For the variation of the dark matter flux we have also considered the range corresponding to the standard Solar System  values $j_\chi=10^{-18}\times 2.5\times 10^7=2.5\times 10^{-11}\;{\rm g/cm^2\;s}$ and an upper limit of $j_\chi=10^{-3}\;{\rm g/cm^2\;s}$.

For the specific case of Planet 9 for the surface temperature of the planet due to kinetic dark matter heating we find the value
\begin{eqnarray}
T_{S9}=16.201&\times& \left(\frac{M}{10M_{\oplus}}\right) ^{1/4}\times \left( \frac{%
R}{10R_{\oplus}}\right) ^{-1/4}\times \left( \frac{\rho _{\chi }}{%
10^{-18}\;{\rm g/cm^{3}}}\right) ^{1/4}\nonumber\\
&&\times \left( \frac{v_{\chi }}{2.50\times 10^7\;{\rm cm/s}}%
\right) ^{1/4}\times \left( \frac{f}{1}\right) ^{1/4}\;{\rm K}.
\end{eqnarray}
 In the static case, the surface temperature essentially depends on the numerical value of the capture rate $f$. For $f=10^{-8}$, the contribution of the kinetic heating by dark matter of the surface temperature of Planet 9 becomes negligibly small.   

\subsubsection{Dynamical temperature evolution}

Assuming now a dynamical time variation of the planet mass, under the
assumption of an approximately constant radius, we obtain
\begin{equation}
\frac{dE_{k}(t)}{dt}=\pi GM_{0}R\rho _{\chi }v_{\chi }e^{t/t_{c}}f=4\pi \sigma
_{B}R^{2}T_{S}^{4},
\end{equation}%
giving
\begin{eqnarray}
T_{S}(t)=\left( \frac{2GM_{0}}{R}\frac{1}{8\sigma _{B}}\rho _{\chi }v_{\chi
}e^{t/t_{c}}f\right) ^{1/4}=\left(\frac{1}{8\sigma_B}v_{esc0}^2j_\chi e^{t/t_c}f\right)^{1/4}=T_S^{(0)}e^{t/4t_c},
\end{eqnarray}
where we have denoted $v_{esc0}^2=2GM_0/R$, and $T_S^{(0)}=\left[\left(1/\sigma_B\right)v_{esc0}^2j_\chi f\right]^{1/4}$, respectively.

In order to estimate the time evolution of the surface temperature of Planet 9 we need to estimate for the case of Planet 9 the critical time $t_{c}$, given by Eqs. (\ref{tc1}) and (\ref{tc2}), respectively, which depend on the value of the parameter $\alpha$. We estimate the value of $\alpha$ by using Eq.~(\ref{alpha}), and thus we adopt for it the value $\alpha =1.50\times 10^5$.  Then for the critical time $t_c$ we obtain the expression
\begin{eqnarray}
t_{c9}=6.24\times 10^{16}\times \left(\frac{v_\chi}{2.50\times 10^7\;{\rm cm/s}}\right)\times \left(\frac{R}{10R_\oplus}\right)^{-1}\times \left(\frac{\rho_\chi}{10^{-18}\;{\rm g/cm^3}}\right)^{-1}\;{\rm s}.
\end{eqnarray}

We assume that the initial mass of Planet 9 is $n$ times the mass of the Earth, so that $M_0=nM_\oplus$. As for the initial radius of the planet we still consider it as unchanged by the mass increase.   Then for the time variation of the surface temperature of Planet 9 we find the expression
\begin{eqnarray}\label{Tt}
T_{S9}(t)=15.32&\times&n^{1/4}\times \left( \frac{M_{0}}{M_{\oplus}}\right) ^{1/4}\times
\left( \frac{R}{10R_{\oplus}}\right) ^{-1/4}\times \left( \frac{\rho _{\chi }}{%
10^{-18}\;{\rm g/cm^{3}}}\right) ^{1/4}\nonumber\\
&&\times \left( \frac{v_{\chi }}{2.50\times 10^7\;{\rm cm/s}}%
\right) ^{1/4}\times e^{t/4t_{c9}}\times \left( \frac{f}{1}\right) ^{1/4}\;{\rm K}.
\end{eqnarray}

We consider now a scenario in which Planet 9 formed together with the other planets of the Solar System around 4.5 billion years ago. We further assume that in its first stages of evolution  the main heating mechanisms of the planet were internal heating, or accretion from the interstellar medium. However, these processes may have been exhausted in a the first few billion years of the existence of the planet, and it became very cold, and dark, without any "standard" sources of energy. 

At this moment kinetic dark matter heating through dark matter particle capture may become the dominant heating process, leading to a slow  increase of the temperature of the planet in a time frame of a few billion years.  

\begin{figure}[ht!]
    \centering
    \includegraphics[width=0.75\textwidth]{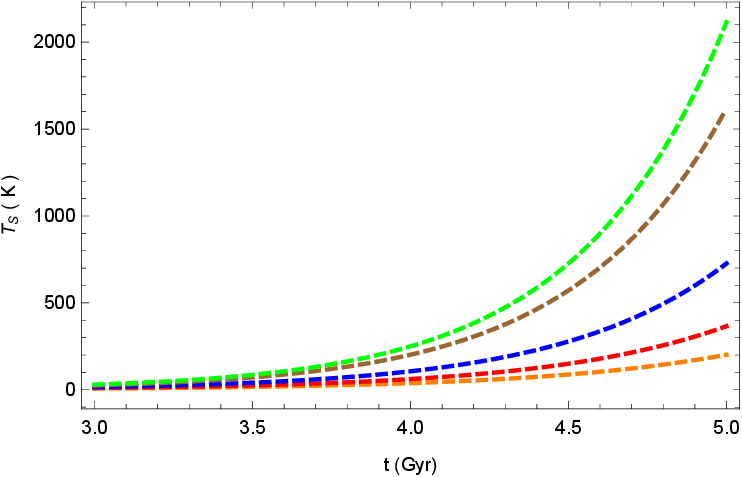}
    \caption{Time Variation of the surface temperature $T_S$ of Planet 9 during a time interval of 4.5 billion years, for an initial planet mass $M_0=M_\oplus$, and for different values of $\rho_\chi$: $\rho_\chi=1.32\times 10^{-17}\;{\rm g/cm^3}$ (orange curve), $\rho_\chi=1.41\times 10^{-17}\;{\rm g/cm^3}$ (red curve), $\rho_\chi=1.52\times 10^{-17}\;{\rm g/cm^3}$ (blue curve),  $\rho_\chi=1.65\times 10^{-17}\;{\rm g/cm^3}$ (brown curve), and $\rho_\chi=1.69\times 10^{-17}\;{\rm g/cm^3}$ (green curve), respectively. For the dark matter capture rate we have assumed the value $f=10^{-10}$.}
    \label{fig2}
\end{figure}

The variation of the surface temperature of Planet 9 is represented, for $f=10^{-10}$, and for different values of the dark matter density and velocity, in Fig.~\ref{fig2}.

As one can see from Fig.~\ref{fig2}, the surface temperature of the planet in the presence of dark matter capture depends extremely sensitively on the environmental dark matter density. Very small changes in $\rho_\chi$ can induce drastic changes on the surface temperature in a long time interval.  This effect is also related to the rapid, exponential increase of the planet mass in the present model. Higher dark matter density environments, with dark matter densities of the order of $\rho_\chi\approx 10^{-15}-10^{-16}\;{\rm g/cm^3}$ could make exoplanets very luminous, and could also provide some important testing methods for the nature of dark matter.  It is also interesting to note that for the present exponential type model, in the first 3 billion years the temperature of the planet remains practically constant, and the effective and efficient heating begins only after a long exposure of the planet to the dark matter flux. 

%By assuming that kinetic dark matter heating was active for a period of 4 %billion years, so that $t=1.26\times 10^{17}$ s, kinetic heating would heat %the surface of Planet 9 to a temperature $\left.T_{S9}(t)\right|_{t=4 \times %10^9 {\rm years}}f^{1/4}=314089\times (f/1)^{1/4}$ K. A period of 3 billions %years of dark matter kinetic heating would give a surface temperature of %$\left.T_{S9}(t)\right|_{t=3 \times 10^9 {\rm years}}f^{1/4}=25490.1\times %(f/1)^{1/4}$ K. 

%A two billions years of heating would lead to a surface temperature %$\left.T_{S9}(t)\right|_{t=2 \times 10^9 {\rm years}}f^{1/4}=2068.66\times %(f/1)^{1/4}$ K, while for one billion years we obtain  $\left.T_{S9}%(t)\right|_{t=1 \times 10^9 {\rm years}}f^{1/4}=167.883\times (f/1)^{1/4}$ K. % A half billion years of kinetic heating would give a surface temperature of %the order of  $\left.T_{S9}(t)\right|_{t=(1/2) \times 10^9 {\rm %years}}f^{1/4}=47.82\times (f/1)^{1/4}$ K. 

\subsection{Spectral signatures of Planet 9}

As we have seen in the previous Section, there is at least a theoretical possibility of the kinetic dark matter heating of Planet 9 to temperatures that make the planet accessible to observations. For simplicity we assume that the dark matter heating can raise, in a period of around 1 billion years, the temperature of the planet to a present day temperature of around 200 K.  Hence Planet 9 appears observationally as a blackbody with a temperature $T_{S9\infty}\sim 200$ K, also depending on the fraction of captured dark matter. According to Wien's displacement law, the maximum of the spectrum of Planet 9 is obtained for $\lambda_{max}T=0.289$, giving for the maximum wavelength the value $\lambda _{max}=1.44\times 10^{-3}$, and a frequency $\nu_{max}=2.07\times 10^{13}\;{\rm s}^{-1}$, which is situated in the infrared domain.

\begin{figure*}[ht!]
    \centering
    \includegraphics[width=0.47\textwidth]{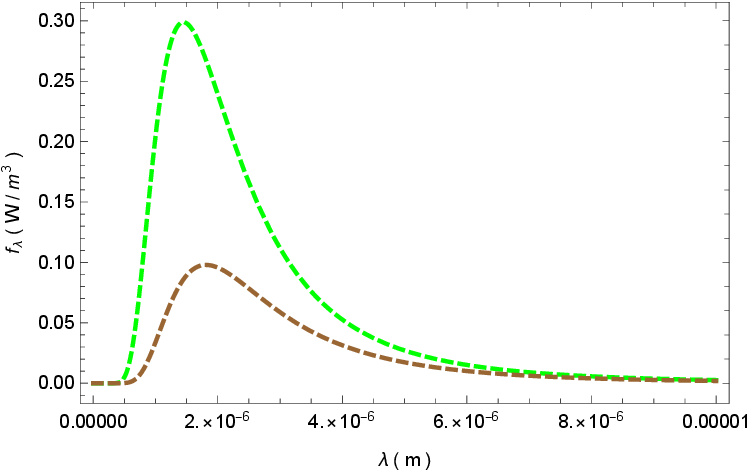}
     \includegraphics[width=0.47\textwidth]{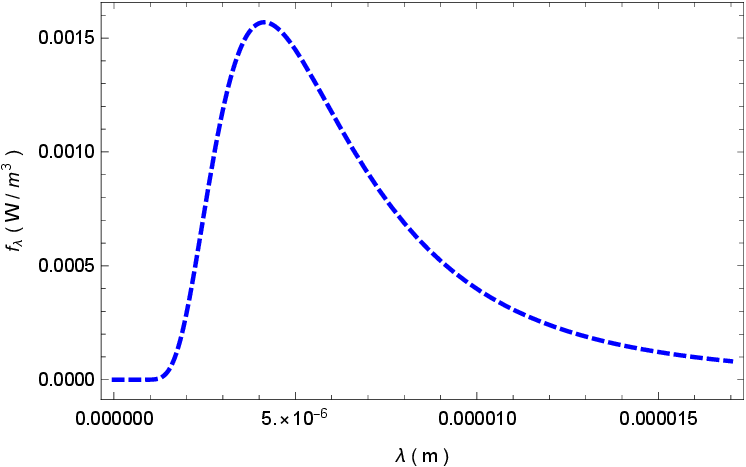}
      \includegraphics[width=0.47\textwidth]{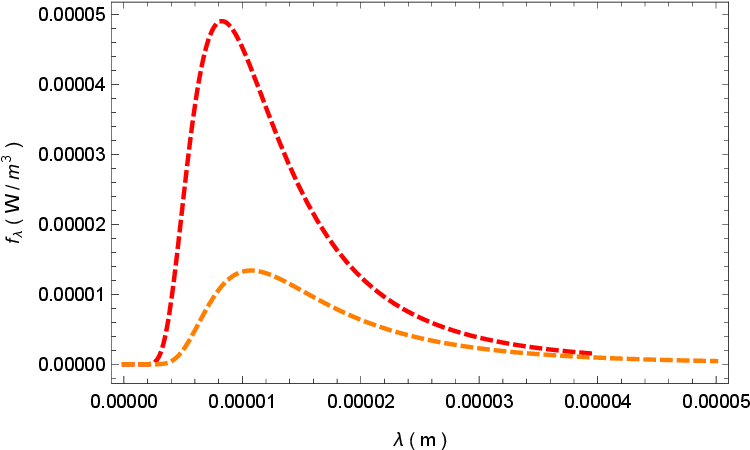}
    \caption{Spectral flux densities $f_\nu$ from the surface of Planet 9 corresponding to different values of $\rho_\chi$ and surface temperature $T_S$: $\rho_\chi=1.69\times 10^{-17}\;{\rm g/cm^3}$, $T_S=2000$ K (green curve), $\rho_\chi=1.65\times 10^{-17}\;{\rm g/cm^3}$, $T_S=1600$ K (brown curve), $\rho_\chi=1.52\times 10^{-17}\;{\rm g/cm^3}$, $T_S=700$ K (blue curve), $\rho_\chi=1.41\times 10^{-17}\;{\rm g/cm^3}$, $T_S=350$ K,  (red curve), and $\rho_\chi=1.32\times 10^{-17}\;{\rm g/cm^3}$, $T_S=250$ K (orange curve), respectively. For the dark matter capture rate, the distance $d$ to Planet 9 and the planet radius we have assumed the values $f=10^{-10}$, $d=500$ UA, and $R=10R_\oplus$, respectively. }
    \label{fig3}
\end{figure*}

The blackbody spectral flux density $f_\nu$ of Planet 9 is given by
\begin{equation}
f_\nu (\nu, T,R,d) =\pi B_\nu (\nu,T)\frac{4\pi R^2}{4\pi d^2},
\end{equation}
where $B_\nu(\nu,T)=\left(2h\nu^3/c^2\right)/\left(e^{h \nu/k_BT}-1\right)$, and $d$ is the distance to the planet,  $k_B=1.38\times 10^{-16} $ ergs/K is Boltzmann's constant, and $h=6.626\times 10^{-27}$ erg s is Planck's constant, respectively. The spectral flux densities corresponding to different dark matter densities, and surface temperatures,  are represented in Fig.~\ref{fig3}.

By assuming that Planet 9 is located at a distance of 500 AU, and has a surface temperature of around $250$ K, for a wavelength  of around $\lambda\approx 10^{-5}$, we obtain $f_{\lambda} \approx 5\times 10^{-5}\;{\rm W/m^3}$. It remains an observational problem if such a low radiation flux density is detectable by the present or future astronomical instruments. 
This means
that if this possibility may exist, and if Planet 9 could be detected observationally,  it could be used to trace the dark matter density distribution in the outer Solar System. Moreover, the overdensities in the dark matter distribution may also be detected, thus obtaining a better understanding on the interaction between dark and normal matter. 

\section{Discussions and final remarks}\label{sec4}

 The basic physical characteristics of Planet 9 are not yet known. Is it a rocky, icy planet, or a giant sphere of gas flowing in space? Or, perhaps, it may be a dark matter dominated object, a so-called dark exoplanet, a type of celestial structures whose existence was suggested in \cite{Dex} (see also \cite{Dex1}) for a discussion of the properties of the dark exoplanets). Dark exoplanets are made of of particles beyond the Standard Model, and especially dark matter particles. If they exist, dark exoplanets present several specific observational signatures, including  including the mass-radius relation, spectroscopy, missing transit, and transit
light curve \cite{Dex}. Two possible transiting exoplanets, assumed to present dark exoplanet signatures have been analyzed in \cite{Dex}. They are CoRoT-1 b \cite{Cor1} and K2-44 b \cite{Cor2}, with masses of the order of  $390.89M_{\oplus}$ (CoRoT-1 b) and $6.5 M_{\oplus}$ (K2-44 b), respectively. Hence, dark exoplanets can have a very wide range of mass distribution, with Planet 9 having a mass close to the mass of K2-44 b. Moreover, the presence of overdensities in the dense dark matter halo can accelerate the increase in mass of the initial proto-planetary seeds, thus making the formation of planetary objects with masses of around $10M_{\oplus}$ possible.

In the present paper we have investigated the possibility of a dark matter heating of Planet 9, a hypothetical object still located in the Solar System, and whose presence could explain a number of gravitational anomalies, including the clustered orbits of distant Kuiper Belt Objects. The tilts and alignments of these objects suggest gravitational interaction  by a massive planet with a mass of about 10 times Earth's mass, and located far beyond Neptune's orbit \cite{Brown1, Brown2,Brown3,Brown4}. 

The possible migration of Planet 9 is the basic approach in explaining its existence at a distance of around 400 to 800 times the Earth-Sun distance \cite{Brown3}. Since not enough material existed in the early outer Solar System to form a massive planet,  Planet 9, having a mass of 5–10 times the mass of Earth, may have formed closer to the Sun, and then it migrated outward towards the boundary of the Solar System. Another possibility is that in fact Planet 9 was captured by the Solar System \cite{Brown3}. However, the migration scenario, which is presently dominant,  should have happened when the Solar System was very young, in the early days of its formation. and thus the assumption that Planet 9 is in its present position for several billion years is a realistic assumption. The positioning of Planet 9 in a low-density baryonic matter region, in the presence of dark matter overdensities,  may have contributed to a significant increase in its dark matter content.

Even though potential Planet 9 candidates have been identified in recent infrared surveys \cite{obs7, obs8}, the problem of its existence is still an open problem, which hopefully will be solved by future observations, performed, for example, by the Vera C. Rubin Observatory \cite{VR1,VR2}.  Still even the significant improvement of observational techniques would require clear observational signatures that could discriminate Planet 9 from other celestial objects. The equilibrium surface temperature of Planet 9 due to Solar luminosity can be approximated as $T_{S9}\approx 54.8\sqrt{26/a_9}$, which gives a value $T_{S9}\approx 13$ K \cite{Chan}. 

An interesting heating mechanism of Planet 9 was considered in \cite{Chan}, where it was suggested that the tidal effect can heat up the satellites of Planet 9 significantly, which can give sufficient thermal radio flux for observations. The tidal heating of the satellites of Planet 9 can be expressed as $T=\left(\dot{E}/4\pi \sigma R^2\right)$, with $\dot{E}=(21C/2)\left(Rn\right)^5e_s^2/G$, where $n=\sqrt{GM_9/a_s^3}$, and $e_s$ is the orbit eccentricity \cite{Chan}.  The tidal heating gives and expected radio flux density $S_\nu \sim 2 \;\mu $Jy for $\nu$ = 300 GHz. However, thee existence of this process requires the existence of satellites of Planet 9.

On the other hand, in the present approach a very long heating period by dark matter of Planet 9 was assumed. If Planet 9 was formed around 4.5 billion years ago, in the first 1-2 billions years of existence the planet could have cleaned its environment from ordinary matter (cosmic dust) by accretion, exhausted its internal energy sources, and became a cold, icy and dark world. At this moment dark matter heating of the planet took over as the dominant physical mechanism, leading, on a long period of several billion years, to the possibility of a significant heating of the surface of Planet 9. Capture of dark matter leads to an increase of the mass of the planet, which we have described by using a simple phenomenological model, giving an exponential time variation of the mass. 

The time evolution of the planet is determined by the critical time $t_c$, which depends on the dark matter parameters, the initial baryonic mass of the planet, and its radius. In the present investigation we have assumed a constant radius for the planet, and a linear mass - radius relation for the planet, an approximation that allows a simpler estimation of the physical parameters of the planet. 

 The theoretical mass-radius relations can be obtained from planetary interior models, once  the equations of state, which relate the density, pressure and temperature is known. In many investigations Small terrestrial type planets are assumed to have constant densities, and thus their mass-radius relation is $R \propto M^{1/3}$ \cite{MR}. However, In massive giant planets with a composition dominated by hydrogen and helium, the gravitational pressure is high enough to make the electron degeneracy pressure to become important. This leads  to a decrease of the radius  with increasing mass, so that $R\propto M^{-1/3}$. The planetary radius also  depends on other factors like stellar irradiation or the planetary age \cite{MR}. On the other hand, a mass-radius relation of the form $R_p\propto M_p^{0.67}$ was obtained in \cite{MR}, using exoplanetary data from the PlanetS catalog, which accounts only for planets with reliable mass and radius determination. Thus a relation is valid for intermediate mass planets, with masses in the range of 4.4 and 127$M_{\oplus}$, with the coefficient of proportionality of the order of 0.56 ($R_p$ and $M_p$ are expressed in Earth units). On the hand, it was found that the radii of giant planets are nearly independent of their masses, $R\propto M^{-0.06}$. 

In the present investigation we have assumed in the numerical investigations a linear mass-radius relation for the Planet 9, an approximation that allows a simpler estimation of the physical parameters of the planet. This is also equivalent in replacing the coefficient 0.67 by 1 in the mass - radius relation. This approximation does not modify significantly the numerical values of the considered physical quantities.

 Assuming a general mass-radius relation of the form $R_p=CM_p^\beta$, where $C$ and $\beta $ are constants, the time variation of the radius of the planet is given by
\begin{equation}
R(t)=CR_\oplus \left(\frac{M_0}{M_\oplus}\right)^{\beta}\left[1+\alpha m(t)\right]^\beta.
\end{equation}

Then the mass capture equation (\ref{mass}) becomes 
\begin{equation}
\frac{dm(t)}{dt}=2\pi G M_0 CR_\oplus \left(\frac{M_0}{M_\oplus}\right)^{\beta}\left[1+\alpha m(t)\right]^{1+\beta},
\end{equation}
with the general solution given by
\begin{equation}
m(t)=\frac{1}{\alpha}\left\{\left[1-2\pi \alpha \beta CG M_0 R_\oplus \left(\frac{M_0}{M_\oplus}\right)^{\beta} t\right]^{-1/\beta}-1\right\}.
\end{equation}
For the variation of the total mass of the planet we find
\begin{equation}
M(t)=M_0\left[1-2\pi \alpha \beta CG M_0 R_\oplus \left(\frac{M_0}{M_\oplus}\right)^{\beta} t\right]^{-1/\beta}.
\end{equation}
The implications of this law of mass variation on the Planet 9 heating will be considered in a future work.

  Due to their interactions with ordinary baryonic matter, non-annihilating dark matter particles can efficiently accumulate inside celestial objects \cite{t2}. High mass dark matter particles can gravitate toward the core of the celestial objects, and thermalize in a small core region.  Stringent constraints on strongly-interacting heavy dark matter can be obtained from the existence of various celestial objects, which can act   optimal detectors to probe the strongly-interacting heavy non-annihilating dark matter \cite{t2}. Generally, the estimates of the dark matter particle mass also depend on the dark matter - neutron cross section, and on the nature of dark matter (bosonic or fermionic). The exclusion limits on spin-independent dark matter-nucleon scattering cross-sections from the existence of several stellar objects presented in \cite{t2} give for the mass of the dark matter particle the approximate range $m_\chi \in \left(10^6,10^{10}\right)$ GeV.

All the present theoretical estimations on the astrophysical parameters of Planet 9 depend on the numerical value of the function $f$, giving the fraction of dark matter particles trapped by the planet. The rate of capture of dark matter was mostly discussed for stars  \cite{star1,star2, star3}. A capture rate of about $f=10^{-10}$ was suggested in \cite{star2,star3}, which in itself is related to many uncertainties in the dark matter capture process. 

By assuming this value for $f$, in the most optimistic scenario of the heating of Planet 9 over a period of 4-5 billion years, the surface temperature of the planet should be of the order of $T_{S9}\approx 1000-2000$ K, while a heating period of 3 billion years would lead to a surface temperature of $T_{S9}\approx 100-200$ K. Dark matter heating over periods of the order of two billion years or less does not give any significant contribution to the surface temperature of the planet in the framework of the present model. However, one must point out that all the estimates of the dark matter heating of Planet 9 depend on the environmental dark matter density, and very small variations in density could lead to significant variations in the surface temperature.   

The existence, properties and orbit of Planet 9 remain a fascinating but very complex and difficult subject of study. Hopefully future new observational data as well as theoretical developments in the field of planetary properties, will decisively contribute in solving the mystery of the ninth planet of the Solar System.

\section*{Acknowledgments}

I would like to thank the two anonymous referees for comments and suggestions that helped me to significantly improve the manuscript. 

\section*{Declarations}

%Some journals require declarations to be submitted in a standardised %format. Please check the Instructions for Authors of the journal to which %you are submitting to see if you need to complete this section. If yes, %your manuscript must contain the following sections under the heading %`Declarations':

\begin{itemize}
\item Funding No funding has been received for this research.
\item Conflict of interest/Competing interests Not applicable.
\item Ethical approval Not applicable.
\item Consent to participate Not applicable.
\item Consent for publication Not applicable.
\item Data availability  All data generated or analysed during this study are included in this published article.
%\item Materials Availability All materials are freely available in the %manuscript
\item Code availability Not applicable. 
%\item Author contribution All authors contributed equally to this work
\end{itemize}

%\noindent
%If any of the sections are not relevant to your manuscript, please %include the heading and write `Not applicable' for that section. 

%%===================================================%%
%% For presentation purpose, we have included        %%
%% \bigskip command. Please ignore this.             %%
%%===================================================%%

%%=============================================%%
%% For submissions to Nature Portfolio Journals %%
%% please use the heading ``Extended Data''.   %%
%%=============================================%%

%%=============================================================%%
%% Sample for another appendix section			       %%
%%=============================================================%%

%% \section{Example of another appendix section}\label{secA2}%
%% Appendices may be used for helpful, supporting or essential material that would otherwise 
%% clutter, break up or be distracting to the text. Appendices can consist of sections, figures, 
%% tables and equations etc.

%\end{appendices}

%%===========================================================================================%%
%% If you are submitting to one of the Nature Portfolio journals, using the eJP submission   %%
%% system, please include the references within the manuscript file itself. You may do this  %%
%% by copying the reference list from your .bbl file, paste it into the main manuscript .tex %%
%% file, and delete the associated \verb+\bibliography+ commands.                            %%
%%===========================================================================================%%

\bibliography{sn-bibliography}% common bib file
%% Journal article

%% if required, the content of .bbl file can be included here once bbl is generated
%\input sn-article.bbl

\end{document}